\documentclass[%
reprint,
 amsmath,amssymb,
 aps,
prl,
]{revtex4-1}

\usepackage{graphicx}% Include figure files
\usepackage{dcolumn}% Align table columns on decimal point
\usepackage{bm}% bold math
\begin{document}
 
\preprint{APS/123-QED}

\title{Many-Anyons Wavefunction, State Capacity and Gentile Statistics}% Force line breaks with \\

\author{Qiang Zhang}
 \email{zhan1217@purdue.edu}
\author{Bin Yan}
\affiliation{%
 Department of Physics and Astronomy, Purdue University, Indiana 47907-2036 
}%

\date{\today}% It is always \today, today,

\begin{abstract}
The many-anyons wavefunction is constructed via the superposition of all the permutations on the direct product of single anyon states and its interchange properties are examined. The phase of permutation is not a representation but the \textit{word metric} of the permutation group . Amazingly the interchange phase yields a finite \textit{capacity} of one quantum state interpolating between Fermion and Boson and the mutual exchange phase has no explicit effect on statistics. Finite capacity of quantum state is defined as \textit{Gentile statistics} and it is different from the fractional exclusion statistics. Some discussion on the general model is also given.
\begin{description}
\item[PACS numbers]
05.30.Pr, 05.30.-d, 05.70.Ce, 03.65.Vf
\end{description}
\end{abstract}

\pacs{Valid PACS appear here}% PACS, the Physics and Astronomy
%Use showkeys class option if keyword
\maketitle

%\tableofcontents
%\section{Introduction}
In the past three decades, the topic of anyon\cite{wilczek90} and fractional statistics is continually of great interest for physics community\cite{laughlin88,nayak08}. Anyon is defined through the interchange phase\cite{wilczek82,wu84,halperin84,arovas84,leinaas77}: quasi-particles with arbitrary interchange phase $\theta$ could exist and lead to an intermediate statistics\cite{wilczek82} interpolating between Boson($\theta=0$) and Fermion($\theta=\pi$). The widely analyzed model is the charge-flux model\cite{wilczek82,wu84m} and spin lattice\cite{hatsugai96,kitaev06,Han07}, although the interchange property for many-body wavefunction is never carefully examined. Alternatively, in condensed matter system the one-particle Hilbert space dimension $d_\alpha$ in general depends on the occupation of state, $\Delta d_\alpha=-\sum_\beta g_{\alpha\beta}\Delta N_\beta$, yielding fractional exclusion statistics proposed by Haldane-Wu\cite{haldane91,wu94}. The statistical parameter $g_{\alpha\beta}$ might relate to the interchange phase, e.g. $g_{\alpha\beta}=\theta_{\alpha\beta}/(2\pi)$ in FQHE\cite{haldane91} and $g_{\alpha\beta}=\delta_{\alpha\beta}\theta/\pi $ in spin chain\cite{wu94,hatsugai96} with possible verification\cite{cooper15,trovato13}.  Yet, an universal relation between the interchange phase and statistics effect is still lacking. The aim of this paper is to construct the many-anyons wavefunction and explore its statistics.\\

%\section{Many-anyons wavefunction}\label{many}
\textit{Many-anyons wavefunction}-The physical interchange of anyons may give any(rational) phase\cite{wilczek82}, $\phi^\dagger_a\phi^\dagger_b=e^{i\theta}\phi^\dagger_b\phi^\dagger_a$, specifically in two dimension as a representation of the braid group\cite{wu84,leinaas77}. Analogy to the Slater determinant form of many-Fermions wavefunction, we could build up our many-anyons wavefunction(unnormalized) by summing over all the permutations on the direct product of single particle states: 
\begin{align}\label{psi}
\Psi\equiv\sum_P(e^{-i\theta})^{l(P)}P\Phi_0
\end{align}
here $\Phi_0 $ is the direct product of the orthonormal single anyon states $\phi_a$ with a preset order, let's say $\Phi_0=\phi_a\bigotimes\phi_b\cdots$(omit $\bigotimes $ for later convenience). $P$ belongs to the permutation group $S_N$. $P\Phi_0$ is a permutation of the ordered anyons and its coefficient is a phase $e^{-i\theta l(P)}$ to guarantee their equal probability. Once we select the phase of $\Phi_0$ as zero, then we will expect the phase of $P\Phi_0$ with once adjacent interchange to be $\theta$. Thus, it is reasonable to define $l(P)$ as the number of \textit{least adjacent interchange} from order $\Phi_0 $ to $P\Phi_0 $. This is consistent with the field operator language where we are barely allowed to interchange neighbor operators one by one. $l(P)$ is called \textit{length} of $P$ in \textit{word metric}\cite{billey07} and it equals to the number of \textit{inversion pairs} (nm) in ordered $P\Phi_0$(anyon-$m$ appears on the left of anyon-$n$ in $P\Phi_0$ for $m>n$)\cite{BH01}. Typically, $l(PP^\prime)\leq l(P)+l(P^\prime)$ and surely $e^{-i\theta l(P)}$ is \textit{not} a representation of the permutation group, unless $e^{-i\theta}=\pm1$. The negative sign in the phase is to ensure the positive interchange phase of the total wavefunction.\\
To gain more reliability, we label the anyons with upper indices and let $P$ permute on anyons, equivalent to on the states, to check the interchange properties of the wavefunction. $l(P)$ now is the sum over the inversion pairs parameter $g_{nm}$, since those anyons must be interchanged to obtain $P\Phi_0$. For instance, two-anyons
\begin{equation}\label{2}
\Psi_{ab}^{12}\equiv\sum_P e^{-i\theta l(P)}P\phi_a^i\phi_b^j= \phi_a^1\phi_b^2+e^{-i\theta g_{12}}\phi_a^2\phi_b^1
\end{equation}
and three-anyons
\begin{align}\label{3}
\Psi_{abc}^{123}
=& \phi_a^1\phi_b^2\phi_c^3+e^{-i\theta g_{12}}\phi_a^2\phi_b^1\phi_c^3+e^{-i\theta g_{12}}e^{-i\theta g_{13}}\phi_a^2\phi_b^3\phi_c^1\nonumber\\
 & +e^{-i\theta g_{12}}e^{-i\theta g_{13}}e^{-i\theta g_{23}}\phi_a^3\phi_b^2\phi_c^1+e^{-i\theta g_{23}} \phi_a^1\phi_b^3\phi_c^2\nonumber\\
&+e^{-i\theta g_{23}}e^{-i\theta g_{13}} \phi_a^3\phi_b^1\phi_c^2
\end{align}
For two-anyons wavefunction, interchange anyon 1 and 2 denoting as $\hat{T}^{12}$,
\begin{align}
\hat{T}^{12}\Psi_{ab}^{12}=&\phi_a^2\phi_b^1+e^{-i\theta g_{21}}\phi_b^2\phi_a^1\nonumber\\
=&e^{-i\theta g_{21}}(\phi_a^1\phi_b^2+e^{i\theta g_{21}}\phi_b^1\phi_a^2)
\end{align}
If $g_{21}=-g_{12}$, then the interchange yields a phase shift $\hat{T}^{12}\Psi^{12}_{ab}=e^{-i\theta g_{21}}\Psi^{12}_{ab}$. Certainly interchange twice 
\begin{align}
\hat{T}^{12}(\hat{T}^{12}\Psi^{12}_{ab})=e^{-i\theta g_{12}}e^{-i\theta g_{21}}\Psi^{12}_{ab}=\Psi^{12}_{ab}
\end{align} 
gives no effect. We shall emphasize here that $\hat{T}^{ij}$ also interchange the indices in the phase term. Let $g_{nm}=-g_{mn}=1$ for $n\leq m$ in general, corresponding to the clockwise/count-clockwise winding. Then the permutation of anyons leads to a total phase shift(see the proof in Appendix.I\ref{app:int}), 
\begin{align}
\hat{T}^{ij}\Psi=&e^{i\theta(2|i-j|-1)}\Psi\label{interchange}\\
P\Psi=&e^{i\theta l(P)}\Psi
\end{align}
For different permutation, the phase shift for the wavefunction is not identical, otherwise it must be a trivial representation of permutation group as Boson. The interchange phase of the wavefunction(\ref{interchange}) is anyons label dependent and this is due to the selection of anyons order in $\Phi_0$. Different selection of $\Phi_0$ gives different wavefunction $\Psi^\prime$ and yields different interchange phases. Yet, this does \textit{not} hurt the indistinguishability and statistics since $\Psi^\ast\Psi=\Psi^{\prime\ast}\Psi^\prime $.\\
In the \textit{totally ordered} coordinate space $X$, the anyon state $\phi_a^i$ is projected as wavefunction $\phi_a(x_i)$. Now the coordinate $x_i$ works as label and $g_{ij}$ is replaced with $g(x_i,x_j)$. $g(x_i,x_j)=-g(x_j,x_i)=1$ for $x_i\preceq x_j$. Here $\preceq$ is the order relation symbol. In the region $x_1\preceq x_2\preceq\cdots \preceq x_N$, all the involved inversion pair parameter $g(x_i,x_j)=1$ and the wavefunction is equation(\ref{psi}). In other regions we could permute the coordinates to increasing order and the wavefunction equals to the permutation on wavefunction(\ref{psi}). Compare with the charge-flux model\cite{wu84m,wilczek90}, our many-anyons wavefunction can ensure the permutation properties for many anyons and work for any Hamiltonian system manifesting the anyonic interchange phase.\\

%\section{Capacity of State}
\textit{Capacity of state}-The many-anyons wavefunction(\ref{psi}) is neither symmetric as Boson nor anti-symmetric as Fermion. Consider two and three anyons at the same state($a=b=c$), equations (\ref{2}) and (\ref{3})becomes 
\begin{align}\label{23same}
\Psi_{aa}=&(1+e^{-i\theta})\phi_a\phi_a\\
\Psi_{aaa}=&(1+e^{-i\theta})(1+e^{-i\theta}+e^{-2i\theta})\phi_a\phi_a\phi_a
\end{align}
By induction(or see the proof through generating function in Appendix.II\ref{app:same}) the wavefunction of $N_a$ anyons at the same state-$a$, 
\begin{align}\label{manysame}
\Psi_{N_a}=&\prod_{n=0}^{N_a-1}(\sum_{k=0}^ne^{-i*k\theta})\times\phi_a...\phi_a\nonumber\\
=& \prod_{n=1}^{N_a}(\frac{1-e^{-i*n\theta}}{1-e^{-i\theta}})\times\phi_a...\phi_a
\end{align} 
Indeed, this is the \textit{Q-factorial} form for the number operator in quantum group\cite{macfarlane89} with $Q=e^{-i\theta}$. Analogy to the exclusion of Fermion, where more than two Fermions at the same state vanish the total wavefunction(\ref{23same}), i.e.$1+e^{-i\pi}=0$, the many-anyons wavefunction(\ref{manysame}) vanishes when $N_a\geqslant q+1$, providing $(e^{-i\theta})^{1+q}=1$. As a consequence, the maximal occupation or \textit{capacity} of state-$a$ is $q$. This is the \textit{general exclusion principle} we obtained from the interchange phase!\\
In order that a thermodynamic limit can be achieved via a sequence of systems with different particle numbers\cite{haldane91,wilczek82}, the phase $\theta/\pi$ shall be a rational number $r/p$. The capacity $q$ of quantum state is always an integer,
  \begin{align}
    q=\left\{
                \begin{array}{ll}
                  2p-1,\ \ r\  is\  odd\\
                  p-1,\ \ \ r\  is \ even\\
                \end{array}
              \right.
  \end{align}
Irrational phase factor $\theta/\pi$ can not vanish the total wavefunction, thus yielding no constraint on the occupation. In the (real)Q-analogy quantum group approach\cite{macfarlane89,fivel90,greenberg91}($aa^\dagger +Qa^\dagger a=Q^N$ with $-1<Q<1$), it is noticed that the number operator is positive definite unless $Q$ is the root of unity. If we could extend their $Q$ to the anyonic interchange phase $e^{-i\theta}$, the capacity of quantum state should be found.\\
Many-anyons wavefunction with mutual exchange phases $\theta_{\alpha\beta} $ among species $\alpha, \ \beta$ could be similarly achieved. Basically we construct the wavefunction via equation(\ref{psi}) with $\theta l(P)$ replaced by $\Theta(P)=\sum \theta_{\sigma_i\sigma_j} $. $(ji)$ is inversion pair in $P\Phi_0$ likewise and $\sigma_i=\alpha,\beta\cdots$ is the specie anyon-$i$ belonging to. For $N_a$ anyons at state-$a$ of specie $\alpha$ and $N_b$ anyons at state-$b$ of specie $\beta$,
\begin{align}\label{mut}
\Psi_{N_a^\alpha N_b^\beta}=&\prod_{n=0}^{N_a-1}(\sum_{k=0}^ne^{-i*k\theta_{\alpha\alpha}})\prod_{n=0}^{N_b-1}(\sum_{k=0}^ne^{-i*k\theta_{\beta\beta}})\nonumber\\&
\times\sum e^{-i\Theta(P)}P\phi_\alpha^{N_a}\phi_\beta^{N_b}
\end{align}
In the last sum of permutation terms, there are $(N_a+N_b)!/(N_a!N_b!)$ distinct permutations and all the mutual exchange phases $\theta_{\alpha\beta} $ appear in these terms. They are mutual orthogonal and the superposition will never cancel each other. Thus the wavefunction can only vanish due to the first and second Q-factorial terms of the same state interchange phase. Consequently, the occupation of the same species different states($\alpha=\beta,\ a\neq b$) and of the different species($\alpha\neq \beta$) do not mutually affect. Only the interchange phase of the same species $\theta_{\alpha\alpha}$ constrain the capacity of each single state, quite different from the exclusion statistics parameter $g_{\alpha\beta}$ from fractional exclusion statistics\cite{haldane91,wu94}.\\ 
As to the generalized ideal gas of fractional exclusion statistics without mutual statistics\cite{wu94} $g_{\alpha\beta}=\alpha\delta_{\alpha\beta} $, specifically, a Fermi-like step distribution is found at $T=0$. Below the Fermi surface, the occupation number is $1/\alpha$. The statistics parameter is mapped to the phase by $\alpha=\theta/\pi=r/p$, so the maximal occupation $\bar{n}_{\epsilon<\varepsilon_F}$ is $p/r$, not necessary an integer. This is not the same to the capacity we find here either. For instance, when $\theta=2\pi/3$, the state capacity is $q=2$ while from fractional exclusion statistics, $\bar{n}_{\epsilon<\varepsilon_F}=3/2$ and for $\theta=\pi/2$, we get $q=3$ while $\bar{n}_{\epsilon<\varepsilon_F}=2$.\\

\begin{figure}[h]
\includegraphics[width=\linewidth]{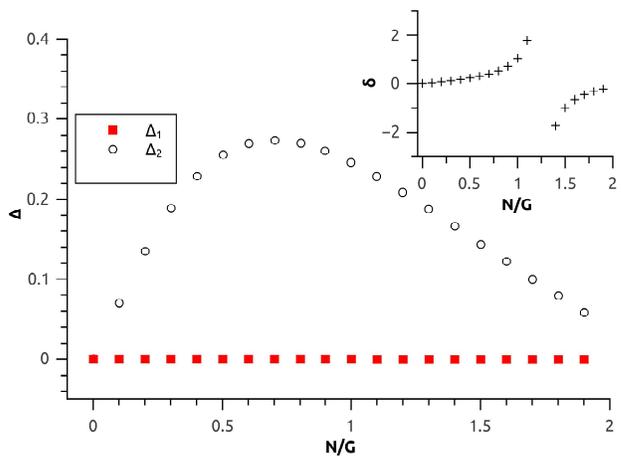}% Here is how to import EPS art
\caption{\label{fig:comp} Numerical comparison of the variance: $\Delta_1\equiv \langle N\rangle/\bar{N}-1$ is the variance between mean occupation and most probable occupation for Gentile statistics. $\Delta_2\equiv \langle N\rangle /\bar{N^\prime}-1$ compares the mean occupation of Gentile statistics and the most probable occupation of fractional exclusion statistics. Here we set $G=10^4$ and $q=3$ as an example. The inset is the relative difference of particle number differentiation $\delta\equiv \delta_N \log W/\delta_N\log W^\prime-1$, which represents the difference of most probable occupation between Gentile and fractional exclusion statistics.}
\end{figure}

%\section{Gentile Statistics}
\textit{Statistics}- From our construction of many-anyons wavefunction, the explicit statistical effect for anyon is that the interchange phase $\theta$ gives a finite capacity of each quantum state. Finite capacity $q$ of quantum state is defined as Gentile statistics\cite{gentile40,dai04,khare05}. The grand canonical partition function is
\begin{align}
Z=&\sum_{\{n_i\}}\sum_iz_i^{n_i}=\prod_i\sum_{n_i=0}^q z_i^{n_i}\\
=&\sum_{\{N_j\}}W(\{N_j\})z_j^{N_j}=\prod_j\sum_{N_j=0}^{qG_j}W_jz_j^{N_j}\label{partition}
\end{align}
here $z_i\equiv e^{-\beta(\epsilon_i-\mu)}$, $\{n_i\}$ denote all the possible configuration of $n_i $ anyons at quantum state-$i$ and $\{N_j\}$ denote all the possible configuration of $N_j$ anyons at energy level $\epsilon_j$ with degeneracy $G_j$. The statistical weight $W(\{N_j\})=\prod_jW_j$ and $W_j$ is thus(\textit{combination with limited repetition}\cite{lavenda91}) 
\begin{equation}\label{we}
W_j=\sum_{k=0}^{k=\lfloor N_j/(q+1)\rfloor}(-1)^kC^k_{G_j}*C^{G_j-1}_{G_j-1+N_j-k(q+1)}
\end{equation}
here Binomial constant $C^k_G\equiv G!/(k!(G-k)!)$. Then the mean occupation is
\begin{align}\label{mean}
\langle n_i\rangle\equiv &\sum_{\{n_i\}}n_i\sum_iz_i^{n_i}=\frac{z_i}{1-z_i}+\frac{(1+q)z^{1+q}}{z_i^{1+q}-1}=\langle N_i\rangle/G_i
\end{align} 
 The thermodynamics behavior such as the heat capacity, equation of state and condensation temperature\cite{lavagno00} are well studied\cite{khare05,dai04}. From equation(\ref{partition}), the most probable occupation $\bar{N}_j$ is determinant from the identity
\begin{align}\label{mp}
 \delta_{N_j} \log (W_jz_j^{N_j})=\delta_{N_j}\log W_j|_{\bar{N}_j}+\log z_j=0
\end{align}
Compare equations (\ref{mean}) and (\ref{mp}), we can estimate the variance between the mean occupation and the most probable occupation $\Delta\equiv \langle N_j\rangle/\bar{N}_j-1$. It can be numerically verified that the variance is almost zero(filled square in FIG.\ref{fig:comp}) coinciding with intuition.
Certainly, at $T=0$, $\langle n_i\rangle=q$ for those states $\epsilon_i<\mu$, similar to the fractional exclusion statistics\cite{haldane91,wu94}. Plausibly, one may consider that $N_j$ particles occupy at least $\lceil N_j/q\rceil$(ceiling function) states among $G_j$ states, then the effective Fock space dimension $d_B=G_j+N_j-\lceil N_j/q\rceil\approx G_j+(N_j-1)(1-1/q)$ and the number of ways $W^\prime_j=C^{N_j}_{d_B}$ is the same to Wu's counting\cite{wu94}, providing $q=1/\alpha$.\\
Yet numerically $W_j\gg W^\prime_j$ in general. From identity(\ref{mp}), the most probable occupation is controled by the differentiation on particle number $\delta_N \log W(N)$. It is close for the two different kinds of counting only in the limit of $G_j\gg N_j$ as shown in the inset of FIG.\ref{fig:comp}. This corresponds to the most probable occupation number of high energy state and both of them reduce to the classical Boltzmann distribution. For the low energy state, $N_j\approx G_j $, Gentile statistics and fractional exclusion statistics yield completely different occupation number(see also empty circle data in FIG.\ref{fig:comp}).\\

\textit{Discussion}-
It is confusing to talk about many anyons at the same state due to the interchange phase, since we would have $\Psi_{aa}=e^{i\theta}\Psi_{aa}$. For $\theta\neq 0$, it seems that the wavefunction vanishes. In some proposed model, such as the magnon excitation in Heisenberg spin chain via Bethe ansatz\cite{hatsugai96,karbach00}, the charge-flux model\cite{wu84m} and the spin-anyon mapping from Jordan-Wigner transformation\cite{fradkin89,batista01}, people always preinstall the hard-core condition, resulting in the conventional Fermi statistics\cite{hatsugai96}. Indeed in the coordinate space $X^N/S_N$\cite{leinaas77} of $N$ indistinguishable particles, the permutation of coordinates at point$(x_1,x_2..x_N)$ gives the identical point while the wavefunction feels a phase shift\cite{leinaas77}. The wavefunction is multivalued\cite{wu84m,leinaas77} and the interchange property($\Psi=e^{i\theta}\Psi$) does not directly indicate a vanishing wavefunction. The hard-core condition is a too strong constraint for identical particles and it makes sense to discuss many-anyons at the same state.\\
We also noticed that the capacity of quantum state is super sensitive to the interchange phase. Two very close phases $\theta$ and $\theta^\prime$ might yield completely different capacities. Thus to obtain a stable capacity and statistics, the interchange phase shall be exact. If the phase factor $r/p$ could be looked as the ratio of electron and flux, thus the resistance, then the resistance must be exactly on the plateau, well known result in FQHE\cite{tsui82,stern08}.\\
In our construction of many-anyons wavefunction(\ref{psi}) from single anyon state, the permutation is decomposed into least adjacent interchange. We applied the winding interpretation $g_{nm}=-g_{mn}=1$ for $n<m$ and (\ref{length}) reduces to the length of $P$ in the word metric. Then anyons shall obey Gentile statistics. Yet, the permutation property(\ref{permute}) is valid for any $g_{nm}=-g_{mn}$, e.g. $g_{nm}=-g_{mn}=(-1)^{n-m-1}$ for $n<m$. The statistics effect may be discussed in future work.\\
 More generally, the many-anyons wavefunction is
\begin{align}\label{anyon}
 \Psi\equiv \sum_Pe^{-i\Theta(P)}P\Phi_0
\end{align}
here $\Phi_0$ could be any function(not necessary to be the product of single anyons state) and $P$ permutes on any indistinguishable indices. The permutation $P^\prime$ on the wavefunction
\begin{align}
P^\prime\Psi=&\sum_P e^{-iP^\prime\Theta(P)}P^\prime P\Phi_0\nonumber\\
=&\sum_P e^{-iP^\prime\Theta(P)+i\Theta(P^\prime P)}e^{-i\Theta(P^\prime P)}P^\prime P\Phi_0\nonumber\\
=&e^{i\Theta^\prime(P^\prime)}\sum_{P}e^{-i\Theta(P^\prime P)}P^\prime P\Phi_0
\end{align}
if
\begin{align}\label{gel}
 \Theta(P^\prime P)-P^\prime\Theta(P)=\Theta^\prime(P^\prime)
\end{align} 
is independent of $P$, then the permutation $P^\prime$ yield a total phase shift $\Theta^\prime(P^\prime)$ for $\Psi$. Equations (\ref{anyon}) and (\ref{gel}) are the general definition for anyons we proposed. There might be other form for $\Theta(P)$ satisfying equation(\ref{gel}) and thus controlling other anyonic statistics.\\
The many-anyons wavefunction equations (\ref{psi}) and (\ref{anyon}) may be used in the consensed matter system and the thermodynamics of Gentile statistics might be compared with experiments.\\
\begin{acknowledgments}

\end{acknowledgments}

\appendix

\section{I.Interchange properties}\label{app:int}
For convenience, we write the ordered anyons sequence $\phi^i_a\phi^j_b\phi^k_c\phi^l_d..$ as $P^{ijkl..}\Phi_0$. Sequence $ijkl..$ could represent the permutation $P$. Any permutation could be decomposed into least adjacent interchange called \textit{reduced word} in word metric\cite{billey07}. Although the decomposition is not unique, the interchanged anyons pairs are identical. They are the \textit{inversion} pairs $(mn)$ in sequence $ijkl..$, if $m$ is on the left of $n$ for $m>n$\cite{BH01}. In terms of inversion pair parameter $g_{nm}$
\begin{align}\label{length}
l(P^{ijkl..})=\sum_{(mn)} g_{nm}
\end{align}
Now we will prove the interchange property of the many-anyons wavefunction $\Psi^{1234\cdots}_{abcd\cdots} $. Consider the following two sequences($i>j$): 
\begin{align}
\circledS1=\underbrace{a_1a_2...a_{k-1}}_{s1}\ i\ \underbrace{a_{k+1}a_{k+2}..a_{m-1}}_{s2}\ j\ \underbrace{a_{m+1}a_{m+2}..a_n}_{s3}\\
\circledS2= \underbrace{a_1a_2...a_{k-1}}_{s1}\ j\ \underbrace{a_{k+1}a_{k+2}..a_{m-1}}_{s2}\ i\ \underbrace{a_{m+1}a_{m+2}..a_n}_{s3}
\end{align} 
 
\begin{align}
l(P^{\circledS1})=&\sum_{m_1>j}g_{jm_1}+\sum_{m_1>i}g_{im_1}+\sum_{m_2<j}g_{m_2j}+\sum_{m_2>i}g_{im_2}\nonumber\\
&+\sum_{m_3<j}g_{m_3j}+\sum_{m_3<i}g_{m_3i}+\textit{ij-indept terms.}\\
l(P^{\circledS2})=&\sum_{m_1>i}g_{im_1}+\sum_{m_1>j}g_{jm_1}+\sum_{m_2<i}g_{m_2i}+\sum_{m_2>j}g_{jm_2}\nonumber\\
&+\sum_{m_3<i}g_{m_3i}+\sum_{m_3<j}g_{m_3j}+g_{ji}+\textit{ji-indept}.
\end{align}
$m_\alpha$($\alpha=1,2,3$) are the anyons in segment $s_\alpha$ of the sequences.
Interchange anyon-$i$ and $j$ by $\hat{T}^{ij}$, sequences $\circledS2$ and $\circledS1$ are mutually mapped. The phase shift shall be equal,
\begin{align}
l_{ij}\equiv\hat{T}^{ij}l(P^{\circledS2})-l(P^{\circledS1})=\hat{T}^{ij}l(P^{\circledS1})-l(P^{\circledS2})
\end{align}
The above identity shall be independent of the position of $i,j$ and the order of other anyons. The only solution for the above equation is $g_{mn}=-g_{nm}$ and the overall phase shift is $e^{-i\theta l_{ij}}$, 
\begin{align}
l_{ij}=\sum_{m>j}^{m<i}(g_{mj}+g_{im})+g_{ij}
\end{align}
$\hat{T}^{ij}$ is a 2-\textit{circle} permutation as an element of the permutation group and it is easy to check $l(\hat{T}^{ij})=-l_{ij}$. Thus $\hat{T}^{ij}\Psi= e^{i\theta l(\hat{T}^{ij})}\Psi$. Any permutation could be decomposed as some independent cyclic permutation $P=T^{ijk\cdots}T^{mn\cdots}$. Use the same method, we can prove that for any $P$
\begin{align}\label{permute}
P\Psi=e^{i\theta l(P)}\Psi
\end{align}
 Let $g_{nm}=-g_{mn}=1$ for $n<m$ as the clockwise/cout-clockwise interpretation, then $l(P)$ is the length of $P$ and the wavefunction reduces to equation(\ref{psi}). Replace $\theta l(P)$ with $\Theta(P)=\sum \theta_{\sigma_i\sigma_j} $, the many-anyons wavefunction with mutual interchange phase $\theta_{\alpha\beta}$ can be obtained in the same spirit.
 
\subsection{II.Occupying the same states}\label{app:same}
For $N$ anyons sequences, denote $I_N(k)$ the number of permutation sequences with $k\leq C^2_N$ inverse pairs. $C_N^m\equiv N!/((N-m)!m!)$ is the binomial constant. Clearly, $I_N(0)=1$ and the generating function $\Psi_N(x)$ of $I_N(k)$ satisfying the recurrence relation\cite{BH01}
\begin{align}
\Psi_N(x)&\equiv\sum_{k=0}^{C^2_N}I_N(k)x^k\\
&=(1+x+x^2+\cdots+x^{N-1})\Psi_{N-1}(x)\label{rec}
\end{align}
For $N_a$-anyons at the same state, equation(\ref{psi})
\begin{align}
\Psi_N&=\sum_Pe^{-i\theta l(P)}P\phi_a...\phi_a \\
&=\sum_{l=0}^{C^l_N}I_N(l)(e^{-i\theta })^l\phi_a...\phi_a
\end{align}
is indeed the generating function of $I_N(k)$ with argument $e^{-i\theta}$. $\Psi_1=1$, use equation(\ref{rec}),
\begin{align}
\Psi_{N_a}=&\prod_{n=0}^{N_a-1}(\sum_{k=0}^ne^{-i*k\theta})\times\phi_a...\phi_a\nonumber\\
=& \prod_{n=1}^{N_a}(\frac{1-e^{-i*n\theta}}{1-e^{-i\theta}})\times\phi_a...\phi_a
\end{align} 
For the wavefunction with different states or mutual interchange phases, the identity equation(\ref{mut}) can be checked out in the same way.
%\nocite{*}

\bibliography{anyon}% Produces the bibliography via BibTeX.

\begin{thebibliography}{31}%
\makeatletter
\providecommand \@ifxundefined [1]{%
 \@ifx{#1\undefined}
}%
\providecommand \@ifnum [1]{%
 \ifnum #1\expandafter \@firstoftwo
 \else \expandafter \@secondoftwo
 \fi
}%
\providecommand \@ifx [1]{%
 \ifx #1\expandafter \@firstoftwo
 \else \expandafter \@secondoftwo
 \fi
}%
\providecommand \natexlab [1]{#1}%
\providecommand \enquote  [1]{``#1''}%
\providecommand \bibnamefont  [1]{#1}%
\providecommand \bibfnamefont [1]{#1}%
\providecommand \citenamefont [1]{#1}%
\providecommand \href@noop [0]{\@secondoftwo}%
\providecommand \href [0]{\begingroup \@sanitize@url \@href}%
\providecommand \@href[1]{\@@startlink{#1}\@@href}%
\providecommand \@@href[1]{\endgroup#1\@@endlink}%
\providecommand \@sanitize@url [0]{\catcode `\\12\catcode `\$12\catcode
  `\&12\catcode `\#12\catcode `\^12\catcode `\_12\catcode `\%12\relax}%
\providecommand \@@startlink[1]{}%
\providecommand \@@endlink[0]{}%
\providecommand \url  [0]{\begingroup\@sanitize@url \@url }%
\providecommand \@url [1]{\endgroup\@href {#1}{\urlprefix }}%
\providecommand \urlprefix  [0]{URL }%
\providecommand \Eprint [0]{\href }%
\providecommand \doibase [0]{http://dx.doi.org/}%
\providecommand \selectlanguage [0]{\@gobble}%
\providecommand \bibinfo  [0]{\@secondoftwo}%
\providecommand \bibfield  [0]{\@secondoftwo}%
\providecommand \translation [1]{[#1]}%
\providecommand \BibitemOpen [0]{}%
\providecommand \bibitemStop [0]{}%
\providecommand \bibitemNoStop [0]{.\EOS\space}%
\providecommand \EOS [0]{\spacefactor3000\relax}%
\providecommand \BibitemShut  [1]{\csname bibitem#1\endcsname}%
\let\auto@bib@innerbib\@empty
%</preamble>
\bibitem [{\citenamefont {Wilczek}(1990)}]{wilczek90}%
  \BibitemOpen
  \bibfield  {author} {\bibinfo {author} {\bibfnamefont {F.}~\bibnamefont
  {Wilczek}},\ }\href@noop {} {\emph {\bibinfo {title} {Fractional statistics
  and anyon superconductivity}}},\ Vol.~\bibinfo {volume} {5}\ (\bibinfo
  {publisher} {World Scientific Singapore},\ \bibinfo {year}
  {1990})\BibitemShut {NoStop}%
\bibitem [{\citenamefont {Laughlin}(1988)}]{laughlin88}%
  \BibitemOpen
  \bibfield  {author} {\bibinfo {author} {\bibfnamefont {R.}~\bibnamefont
  {Laughlin}},\ }\href@noop {} {\bibfield  {journal} {\bibinfo  {journal}
  {Physical review letters}\ }\textbf {\bibinfo {volume} {60}},\ \bibinfo
  {pages} {2677} (\bibinfo {year} {1988})}\BibitemShut {NoStop}%
\bibitem [{\citenamefont {Nayak}\ \emph {et~al.}(2008)\citenamefont {Nayak},
  \citenamefont {Simon}, \citenamefont {Stern}, \citenamefont {Freedman},\ and\
  \citenamefont {Sarma}}]{nayak08}%
  \BibitemOpen
  \bibfield  {author} {\bibinfo {author} {\bibfnamefont {C.}~\bibnamefont
  {Nayak}}, \bibinfo {author} {\bibfnamefont {S.~H.}\ \bibnamefont {Simon}},
  \bibinfo {author} {\bibfnamefont {A.}~\bibnamefont {Stern}}, \bibinfo
  {author} {\bibfnamefont {M.}~\bibnamefont {Freedman}}, \ and\ \bibinfo
  {author} {\bibfnamefont {S.~D.}\ \bibnamefont {Sarma}},\ }\href@noop {}
  {\bibfield  {journal} {\bibinfo  {journal} {Reviews of Modern Physics}\
  }\textbf {\bibinfo {volume} {80}},\ \bibinfo {pages} {1083} (\bibinfo {year}
  {2008})}\BibitemShut {NoStop}%
\bibitem [{\citenamefont {Wilczek}(1982)}]{wilczek82}%
  \BibitemOpen
  \bibfield  {author} {\bibinfo {author} {\bibfnamefont {F.}~\bibnamefont
  {Wilczek}},\ }\href@noop {} {\bibfield  {journal} {\bibinfo  {journal}
  {Physical Review Letters}\ }\textbf {\bibinfo {volume} {49}},\ \bibinfo
  {pages} {957} (\bibinfo {year} {1982})}\BibitemShut {NoStop}%
\bibitem [{\citenamefont {Wu}(1984{\natexlab{a}})}]{wu84}%
  \BibitemOpen
  \bibfield  {author} {\bibinfo {author} {\bibfnamefont {Y.-S.}\ \bibnamefont
  {Wu}},\ }\href@noop {} {\bibfield  {journal} {\bibinfo  {journal} {Physical
  Review Letters}\ }\textbf {\bibinfo {volume} {52}},\ \bibinfo {pages} {2103}
  (\bibinfo {year} {1984}{\natexlab{a}})}\BibitemShut {NoStop}%
\bibitem [{\citenamefont {Halperin}(1984)}]{halperin84}%
  \BibitemOpen
  \bibfield  {author} {\bibinfo {author} {\bibfnamefont {B.~I.}\ \bibnamefont
  {Halperin}},\ }\href@noop {} {\bibfield  {journal} {\bibinfo  {journal}
  {Physical Review Letters}\ }\textbf {\bibinfo {volume} {52}},\ \bibinfo
  {pages} {1583} (\bibinfo {year} {1984})}\BibitemShut {NoStop}%
\bibitem [{\citenamefont {Arovas}\ \emph {et~al.}(1984)\citenamefont {Arovas},
  \citenamefont {Schrieffer},\ and\ \citenamefont {Wilczek}}]{arovas84}%
  \BibitemOpen
  \bibfield  {author} {\bibinfo {author} {\bibfnamefont {D.}~\bibnamefont
  {Arovas}}, \bibinfo {author} {\bibfnamefont {J.~R.}\ \bibnamefont
  {Schrieffer}}, \ and\ \bibinfo {author} {\bibfnamefont {F.}~\bibnamefont
  {Wilczek}},\ }\href@noop {} {\bibfield  {journal} {\bibinfo  {journal}
  {Physical review letters}\ }\textbf {\bibinfo {volume} {53}},\ \bibinfo
  {pages} {722} (\bibinfo {year} {1984})}\BibitemShut {NoStop}%
\bibitem [{\citenamefont {Leinaas}\ and\ \citenamefont
  {Myrheim}(1977)}]{leinaas77}%
  \BibitemOpen
  \bibfield  {author} {\bibinfo {author} {\bibfnamefont {J.~M.}\ \bibnamefont
  {Leinaas}}\ and\ \bibinfo {author} {\bibfnamefont {J.}~\bibnamefont
  {Myrheim}},\ }\href@noop {} {\bibfield  {journal} {\bibinfo  {journal} {Il
  Nuovo Cimento B Series 11}\ }\textbf {\bibinfo {volume} {37}},\ \bibinfo
  {pages} {1} (\bibinfo {year} {1977})}\BibitemShut {NoStop}%
\bibitem [{\citenamefont {Wu}(1984{\natexlab{b}})}]{wu84m}%
  \BibitemOpen
  \bibfield  {author} {\bibinfo {author} {\bibfnamefont {Y.-S.}\ \bibnamefont
  {Wu}},\ }\href@noop {} {\bibfield  {journal} {\bibinfo  {journal} {Physical
  Review Letters}\ }\textbf {\bibinfo {volume} {53}},\ \bibinfo {pages} {111}
  (\bibinfo {year} {1984}{\natexlab{b}})}\BibitemShut {NoStop}%
\bibitem [{\citenamefont {Hatsugai}\ \emph {et~al.}(1996)\citenamefont
  {Hatsugai}, \citenamefont {Kohmoto}, \citenamefont {Koma},\ and\
  \citenamefont {Wu}}]{hatsugai96}%
  \BibitemOpen
  \bibfield  {author} {\bibinfo {author} {\bibfnamefont {Y.}~\bibnamefont
  {Hatsugai}}, \bibinfo {author} {\bibfnamefont {M.}~\bibnamefont {Kohmoto}},
  \bibinfo {author} {\bibfnamefont {T.}~\bibnamefont {Koma}}, \ and\ \bibinfo
  {author} {\bibfnamefont {Y.-S.}\ \bibnamefont {Wu}},\ }\href@noop {}
  {\bibfield  {journal} {\bibinfo  {journal} {Physical review B}\ }\textbf
  {\bibinfo {volume} {54}},\ \bibinfo {pages} {5358} (\bibinfo {year}
  {1996})}\BibitemShut {NoStop}%
\bibitem [{\citenamefont {Kitaev}(2006)}]{kitaev06}%
  \BibitemOpen
  \bibfield  {author} {\bibinfo {author} {\bibfnamefont {A.}~\bibnamefont
  {Kitaev}},\ }\href@noop {} {\bibfield  {journal} {\bibinfo  {journal} {Annals
  of Physics}\ }\textbf {\bibinfo {volume} {321}},\ \bibinfo {pages} {2}
  (\bibinfo {year} {2006})}\BibitemShut {NoStop}%
\bibitem [{\citenamefont {Han}\ \emph {et~al.}(2007)\citenamefont {Han},
  \citenamefont {Raussendorf},\ and\ \citenamefont {Duan}}]{Han07}%
  \BibitemOpen
  \bibfield  {author} {\bibinfo {author} {\bibfnamefont {Y.-J.}\ \bibnamefont
  {Han}}, \bibinfo {author} {\bibfnamefont {R.}~\bibnamefont {Raussendorf}}, \
  and\ \bibinfo {author} {\bibfnamefont {L.-M.}\ \bibnamefont {Duan}},\ }\href
  {\doibase 10.1103/PhysRevLett.98.150404} {\bibfield  {journal} {\bibinfo
  {journal} {Phys. Rev. Lett.}\ }\textbf {\bibinfo {volume} {98}},\ \bibinfo
  {pages} {150404} (\bibinfo {year} {2007})}\BibitemShut {NoStop}%
\bibitem [{\citenamefont {Haldane}(1991)}]{haldane91}%
  \BibitemOpen
  \bibfield  {author} {\bibinfo {author} {\bibfnamefont {F.~D.~M.}\
  \bibnamefont {Haldane}},\ }\href@noop {} {\bibfield  {journal} {\bibinfo
  {journal} {Physical review letters}\ }\textbf {\bibinfo {volume} {67}},\
  \bibinfo {pages} {937} (\bibinfo {year} {1991})}\BibitemShut {NoStop}%
\bibitem [{\citenamefont {Wu}(1994)}]{wu94}%
  \BibitemOpen
  \bibfield  {author} {\bibinfo {author} {\bibfnamefont {Y.-S.}\ \bibnamefont
  {Wu}},\ }\href@noop {} {\bibfield  {journal} {\bibinfo  {journal} {Physical
  review letters}\ }\textbf {\bibinfo {volume} {73}},\ \bibinfo {pages} {922}
  (\bibinfo {year} {1994})}\BibitemShut {NoStop}%
\bibitem [{\citenamefont {Cooper}\ and\ \citenamefont
  {Simon}(2015)}]{cooper15}%
  \BibitemOpen
  \bibfield  {author} {\bibinfo {author} {\bibfnamefont {N.~R.}\ \bibnamefont
  {Cooper}}\ and\ \bibinfo {author} {\bibfnamefont {S.~H.}\ \bibnamefont
  {Simon}},\ }\href {\doibase 10.1103/PhysRevLett.114.106802} {\bibfield
  {journal} {\bibinfo  {journal} {Phys. Rev. Lett.}\ }\textbf {\bibinfo
  {volume} {114}},\ \bibinfo {pages} {106802} (\bibinfo {year}
  {2015})}\BibitemShut {NoStop}%
\bibitem [{\citenamefont {Trovato}\ and\ \citenamefont
  {Reggiani}(2013)}]{trovato13}%
  \BibitemOpen
  \bibfield  {author} {\bibinfo {author} {\bibfnamefont {M.}~\bibnamefont
  {Trovato}}\ and\ \bibinfo {author} {\bibfnamefont {L.}~\bibnamefont
  {Reggiani}},\ }\href {\doibase 10.1103/PhysRevLett.110.020404} {\bibfield
  {journal} {\bibinfo  {journal} {Phys. Rev. Lett.}\ }\textbf {\bibinfo
  {volume} {110}},\ \bibinfo {pages} {020404} (\bibinfo {year}
  {2013})}\BibitemShut {NoStop}%
\bibitem [{\citenamefont {Billey}\ and\ \citenamefont
  {Jones}(2007)}]{billey07}%
  \BibitemOpen
  \bibfield  {author} {\bibinfo {author} {\bibfnamefont {S.~C.}\ \bibnamefont
  {Billey}}\ and\ \bibinfo {author} {\bibfnamefont {B.~C.}\ \bibnamefont
  {Jones}},\ }\href@noop {} {\bibfield  {journal} {\bibinfo  {journal} {Annals
  of Combinatorics}\ }\textbf {\bibinfo {volume} {11}},\ \bibinfo {pages} {285}
  (\bibinfo {year} {2007})}\BibitemShut {NoStop}%
\bibitem [{\citenamefont {Margolius}(2001)}]{BH01}%
  \BibitemOpen
  \bibfield  {author} {\bibinfo {author} {\bibfnamefont {B.~H.}\ \bibnamefont
  {Margolius}},\ }\href@noop {} {\bibfield  {journal} {\bibinfo  {journal}
  {Journal of Integer Sequences}\ }\textbf {\bibinfo {volume} {4}},\ \bibinfo
  {pages} {3} (\bibinfo {year} {2001})}\BibitemShut {NoStop}%
\bibitem [{\citenamefont {Macfarlane}(1989)}]{macfarlane89}%
  \BibitemOpen
  \bibfield  {author} {\bibinfo {author} {\bibfnamefont {A.}~\bibnamefont
  {Macfarlane}},\ }\href@noop {} {\bibfield  {journal} {\bibinfo  {journal}
  {Journal of Physics A: Mathematical and general}\ }\textbf {\bibinfo {volume}
  {22}},\ \bibinfo {pages} {4581} (\bibinfo {year} {1989})}\BibitemShut
  {NoStop}%
\bibitem [{\citenamefont {Fivel}(1990)}]{fivel90}%
  \BibitemOpen
  \bibfield  {author} {\bibinfo {author} {\bibfnamefont {D.~I.}\ \bibnamefont
  {Fivel}},\ }\href@noop {} {\bibfield  {journal} {\bibinfo  {journal}
  {Physical review letters}\ }\textbf {\bibinfo {volume} {65}},\ \bibinfo
  {pages} {3361} (\bibinfo {year} {1990})}\BibitemShut {NoStop}%
\bibitem [{\citenamefont {Greenberg}(1991)}]{greenberg91}%
  \BibitemOpen
  \bibfield  {author} {\bibinfo {author} {\bibfnamefont {O.}~\bibnamefont
  {Greenberg}},\ }\href@noop {} {\bibfield  {journal} {\bibinfo  {journal}
  {Physical Review D}\ }\textbf {\bibinfo {volume} {43}},\ \bibinfo {pages}
  {4111} (\bibinfo {year} {1991})}\BibitemShut {NoStop}%
\bibitem [{\citenamefont {Gentile~j}(1940)}]{gentile40}%
  \BibitemOpen
  \bibfield  {author} {\bibinfo {author} {\bibfnamefont {G.}~\bibnamefont
  {Gentile~j}},\ }\href@noop {} {\bibfield  {journal} {\bibinfo  {journal} {Il
  Nuovo Cimento (1924-1942)}\ }\textbf {\bibinfo {volume} {17}},\ \bibinfo
  {pages} {493} (\bibinfo {year} {1940})}\BibitemShut {NoStop}%
\bibitem [{\citenamefont {Dai}\ and\ \citenamefont {Xie}(2004)}]{dai04}%
  \BibitemOpen
  \bibfield  {author} {\bibinfo {author} {\bibfnamefont {W.-S.}\ \bibnamefont
  {Dai}}\ and\ \bibinfo {author} {\bibfnamefont {M.}~\bibnamefont {Xie}},\
  }\href@noop {} {\bibfield  {journal} {\bibinfo  {journal} {Annals of
  Physics}\ }\textbf {\bibinfo {volume} {309}},\ \bibinfo {pages} {295}
  (\bibinfo {year} {2004})}\BibitemShut {NoStop}%
\bibitem [{\citenamefont {Khare}(2005)}]{khare05}%
  \BibitemOpen
  \bibfield  {author} {\bibinfo {author} {\bibfnamefont {A.}~\bibnamefont
  {Khare}},\ }\href@noop {} {\emph {\bibinfo {title} {Fractional statistics and
  quantum theory}}},\ Vol.~\bibinfo {volume} {2}\ (\bibinfo  {publisher} {World
  Scientific},\ \bibinfo {year} {2005})\BibitemShut {NoStop}%
\bibitem [{\citenamefont {Lavenda}(1991)}]{lavenda91}%
  \BibitemOpen
  \bibfield  {author} {\bibinfo {author} {\bibfnamefont {B.~H.}\ \bibnamefont
  {Lavenda}},\ }\href@noop {} {\emph {\bibinfo {title} {Statistical physics}}}\
  (\bibinfo  {publisher} {Wiley},\ \bibinfo {year} {1991})\BibitemShut
  {NoStop}%
\bibitem [{\citenamefont {Lavagno}\ and\ \citenamefont
  {Swamy}(2000)}]{lavagno00}%
  \BibitemOpen
  \bibfield  {author} {\bibinfo {author} {\bibfnamefont {A.}~\bibnamefont
  {Lavagno}}\ and\ \bibinfo {author} {\bibfnamefont {P.~N.}\ \bibnamefont
  {Swamy}},\ }\href@noop {} {\bibfield  {journal} {\bibinfo  {journal}
  {Physical Review E}\ }\textbf {\bibinfo {volume} {61}},\ \bibinfo {pages}
  {1218} (\bibinfo {year} {2000})}\BibitemShut {NoStop}%
\bibitem [{\citenamefont {Karbach}\ \emph {et~al.}(2000)\citenamefont
  {Karbach}, \citenamefont {Hu},\ and\ \citenamefont {Muller}}]{karbach00}%
  \BibitemOpen
  \bibfield  {author} {\bibinfo {author} {\bibfnamefont {M.}~\bibnamefont
  {Karbach}}, \bibinfo {author} {\bibfnamefont {K.}~\bibnamefont {Hu}}, \ and\
  \bibinfo {author} {\bibfnamefont {G.}~\bibnamefont {Muller}},\ }\href@noop {}
  {\bibfield  {journal} {\bibinfo  {journal} {arXiv preprint cond-mat/0008018}\
  } (\bibinfo {year} {2000})}\BibitemShut {NoStop}%
\bibitem [{\citenamefont {Fradkin}(1989)}]{fradkin89}%
  \BibitemOpen
  \bibfield  {author} {\bibinfo {author} {\bibfnamefont {E.}~\bibnamefont
  {Fradkin}},\ }\href@noop {} {\bibfield  {journal} {\bibinfo  {journal}
  {Physical review letters}\ }\textbf {\bibinfo {volume} {63}},\ \bibinfo
  {pages} {322} (\bibinfo {year} {1989})}\BibitemShut {NoStop}%
\bibitem [{\citenamefont {Batista}\ and\ \citenamefont
  {Ortiz}(2001)}]{batista01}%
  \BibitemOpen
  \bibfield  {author} {\bibinfo {author} {\bibfnamefont {C.}~\bibnamefont
  {Batista}}\ and\ \bibinfo {author} {\bibfnamefont {G.}~\bibnamefont
  {Ortiz}},\ }\href@noop {} {\bibfield  {journal} {\bibinfo  {journal}
  {Physical review letters}\ }\textbf {\bibinfo {volume} {86}},\ \bibinfo
  {pages} {1082} (\bibinfo {year} {2001})}\BibitemShut {NoStop}%
\bibitem [{\citenamefont {Tsui}\ \emph {et~al.}(1982)\citenamefont {Tsui},
  \citenamefont {Stormer},\ and\ \citenamefont {Gossard}}]{tsui82}%
  \BibitemOpen
  \bibfield  {author} {\bibinfo {author} {\bibfnamefont {D.~C.}\ \bibnamefont
  {Tsui}}, \bibinfo {author} {\bibfnamefont {H.~L.}\ \bibnamefont {Stormer}}, \
  and\ \bibinfo {author} {\bibfnamefont {A.~C.}\ \bibnamefont {Gossard}},\
  }\href@noop {} {\bibfield  {journal} {\bibinfo  {journal} {Physical Review
  Letters}\ }\textbf {\bibinfo {volume} {48}},\ \bibinfo {pages} {1559}
  (\bibinfo {year} {1982})}\BibitemShut {NoStop}%
\bibitem [{\citenamefont {Stern}(2008)}]{stern08}%
  \BibitemOpen
  \bibfield  {author} {\bibinfo {author} {\bibfnamefont {A.}~\bibnamefont
  {Stern}},\ }\href@noop {} {\bibfield  {journal} {\bibinfo  {journal} {Annals
  of Physics}\ }\textbf {\bibinfo {volume} {323}},\ \bibinfo {pages} {204}
  (\bibinfo {year} {2008})}\BibitemShut {NoStop}%
\end{thebibliography}%

\end{document}